\documentclass[twocolumn,amsmath,amssymb,superscriptaddress,nofootinbib]{revtex4}
\pdfoutput=1

\usepackage[latin1]{inputenc}
\usepackage[english]{babel}
\usepackage{amssymb}
\usepackage{amsmath}
\usepackage{amsthm}
\usepackage[]{graphicx}
\usepackage[]{subfigure}
\usepackage{tensor}
\usepackage{color}
\usepackage{cancel}
\usepackage{setspace}
\usepackage{fancyhdr}
\usepackage[bookmarks,linktocpage, colorlinks=true, plainpages = false, citecolor = blue,  linkcolor=blue, urlcolor = blue, filecolor = blue]{hyperref}

\begin{document}

\allowdisplaybreaks
\begin{titlepage}

\title{The No-Boundary Proposal as a Path Integral with Robin Boundary Conditions \\ $^{}$}

\author{Alice Di Tucci}
\affiliation{Max--Planck--Institute for Gravitational Physics (Albert--Einstein--Institute), 14476 Potsdam, Germany}
\author{Jean-Luc Lehners}
\affiliation{Max--Planck--Institute for Gravitational Physics (Albert--Einstein--Institute), 14476 Potsdam, Germany}

\begin{abstract}
Realising the no-boundary proposal of Hartle and Hawking as a consistent gravitational path integral has been a long-standing puzzle. In particular, it was demonstrated by Feldbrugge et al. that the sum over all universes starting from zero size results in an unstable saddle point geometry. Here we show that in the context of gravity with a positive cosmological constant, path integrals with a specific family of Robin boundary conditions overcome this problem. These path integrals are manifestly convergent and are approximated by stable Hartle-Hawking saddle point geometries. The price to pay is that the off-shell geometries do not start at zero size. The Robin boundary conditions may be interpreted as an initial state with Euclidean momentum, with the quantum uncertainty shared between initial size and momentum. 
\end{abstract}
\maketitle

\end{titlepage}


 
If quantum theory is universal, and there currently is no reason to think otherwise, then the universe should be describable by a quantum state just like any other system. While its quantum properties might be hidden today they may well have played a crucial role in an early phase of its evolution. An intriguing idea in this context is that a finite universe might have made its appearance out of nothing due to something akin of a quantum tunnelling effect. In fact this idea has a long history dating back all the way to Lema\^{i}tre \cite{Lemaitre:1931zzb}. Quantum cosmology should then be able to describe such a process. The most concrete formulations of this idea go under the names of ``no boundary proposal'' \cite{HH} and ``tunnelling proposal'' \cite{Vilenkin:1982de}. The path integral formulation of these proposals has recently been analysed in the context of Lorentzian quantum cosmology \cite{FLT1,FLT2,FLT3,DL,Lehners:2018eeo}. There, a transition amplitude from a geometry of zero size, i.e. from ``nothing'', to a finite one was evaluated. It was shown that the two proposals are in fact identical in this formulation and that the path integral gives a result analogous to the tunnelling amplitude described in \cite{Vilenkin:1982de}. Perturbations around the background geometry however turn out to be unstable, meaning that larger and larger deviations from homogeneity and isotropy are favoured quantum mechanically. This led to the conclusion that neither the no boundary nor the tunnelling proposal can be appropriate descriptions of the initial conditions of our universe. 

In follow-up works different implementations of the proposals were proposed in order to avoid this negative conclusion. One attempt by Diaz-Dorronsoro et al. was to consider intrinsically complex contours of integration for the lapse function \cite{DiazDorronsoro:2017hti}, also in conjunction with a specification of the initial momentum \cite{DiazDorronsoro:2018wro}. However, these modifications either still included unstable perturbations \cite{FLT3} or led to inconsistencies \cite{FLT4}. Another approach by Vilenkin and Yamada (in the context of the tunnelling proposal) was to modify the boundary conditions for the perturbations to be of Robin type \cite{VY1,VY2}. A final proposition by Halliwell, Hartle and Hertog was to simply abandon the path integral, and concentrate on solutions of the Wheeler-DeWitt equation with desirable properties \cite{Halliwell:2018ejl} (see also \cite{deAlwis:2018sec}). However, the path integral neatly captures quantum interference, and focussing on solutions that cannot be described via a path integral may thus not correctly reproduce central quantum effects. 

In this {\it Letter}, we will combine several of the ideas mentioned above, using Robin boundary conditions in order to impose a condition on a linear combination of the initial size and momentum of the universe. For a family of such conditions, we find that the path integral can be approximated solely by the stable no-boundary saddle points, thus avoiding instabilities and representing a consistent definition of the no-boundary proposal.

 
But first, it seems appropriate to briefly review the negative result that we are trying to overcome. We will work in minisuperspace, where the geometries considered are of the form \cite{Halliwell:1988ik} $ds^2 = - \frac{N^2 }{q} dt^2 + q d \Omega_3^2,$  with $ d \Omega_3^2$ a three sphere of volume $2\pi^2$.  The function $q(t)$ is the squared scale factor, while $N(t)$ is the lapse function. The propagator for the no boundary proposal $G[q_1 , 0]$ describes a transition from a spatial 3-geometry of zero size, $q_0 = 0$, to a later one of size $q_1$. The propagator can be evaluated as a path integral over 4-geometries with a weighting $e^{iS}$ given by the Einstein-Hilbert action with a cosmological constant $\Lambda = 3 H^2$. The steps needed to define the gravitational path integral can be found in \cite{teitelboim,halliwell}, with final result (in constant lapse gauge $\dot{N} = 0$) 
  \begin{equation}
  G[q_1 , 0 ] = \int_{0^+}^\infty dN \, \int_{q(t = 0)= 0}^{q(t = 1) = q_1} \delta q \, e^{i S / \hbar}\,,  \label{prop}
  \end{equation}
where the action takes the form (with $8 \pi G =1$)   
  \begin{equation}
  S \! = \! \! \int \! d^4 x \sqrt{-g} \left( \frac{R}{2}\! - \!\Lambda \right) = 6\pi^2 \! \int_0^1 \! \! dt \, [ - \frac{\dot{q}^2}{4 N} + N( 1 - H^2 q)]\,. \nonumber
  \end{equation}
The integration domain for the lapse is $N \in (0^+ , \infty),$ ensuring that the geometries in the sum have a Lorentzian signature. This makes the path integral a propagator in the sense that it solves the inhomogeneous Wheleer-DeWitt equation $\hat{\mathcal{H}} G[q_1  , 0 ] = - i \delta(q_1)$ where $\hat{\mathcal{H}} $ is the quantum Hamiltonian.
 
It has been shown in \cite{FLT1} that the result of the various integrations to leading order in $\hbar$ is 
 \begin{equation}
 G[q_1 , 0] = e^{- \frac{4\pi^2}{H^2 \hbar} - i \frac{4\pi^2 H}{\hbar} (q_1 - \frac{1}{H^2})^{3/2}} \label{G}
\end{equation}  
where we assumed that $q_1 > 1/H^2$. The negative weighting $- \frac{4\pi^2}{\hbar H^2}$ is characteristic of the tunnelling proposal \cite{Vilenkin:1982de}, whereas the conjectured no-boundary result is \cite{HH}
\begin{equation}
\Psi (q_1) =  e^{+\frac{4\pi^2}{H^2 \hbar}} \cos[  \frac{4 \pi^2 H}{\hbar} (q_1 - \frac{1}{H^2})^{3/2}]\,. \label{hh}
\end{equation}  

\begin{figure}
\includegraphics[scale=0.45]{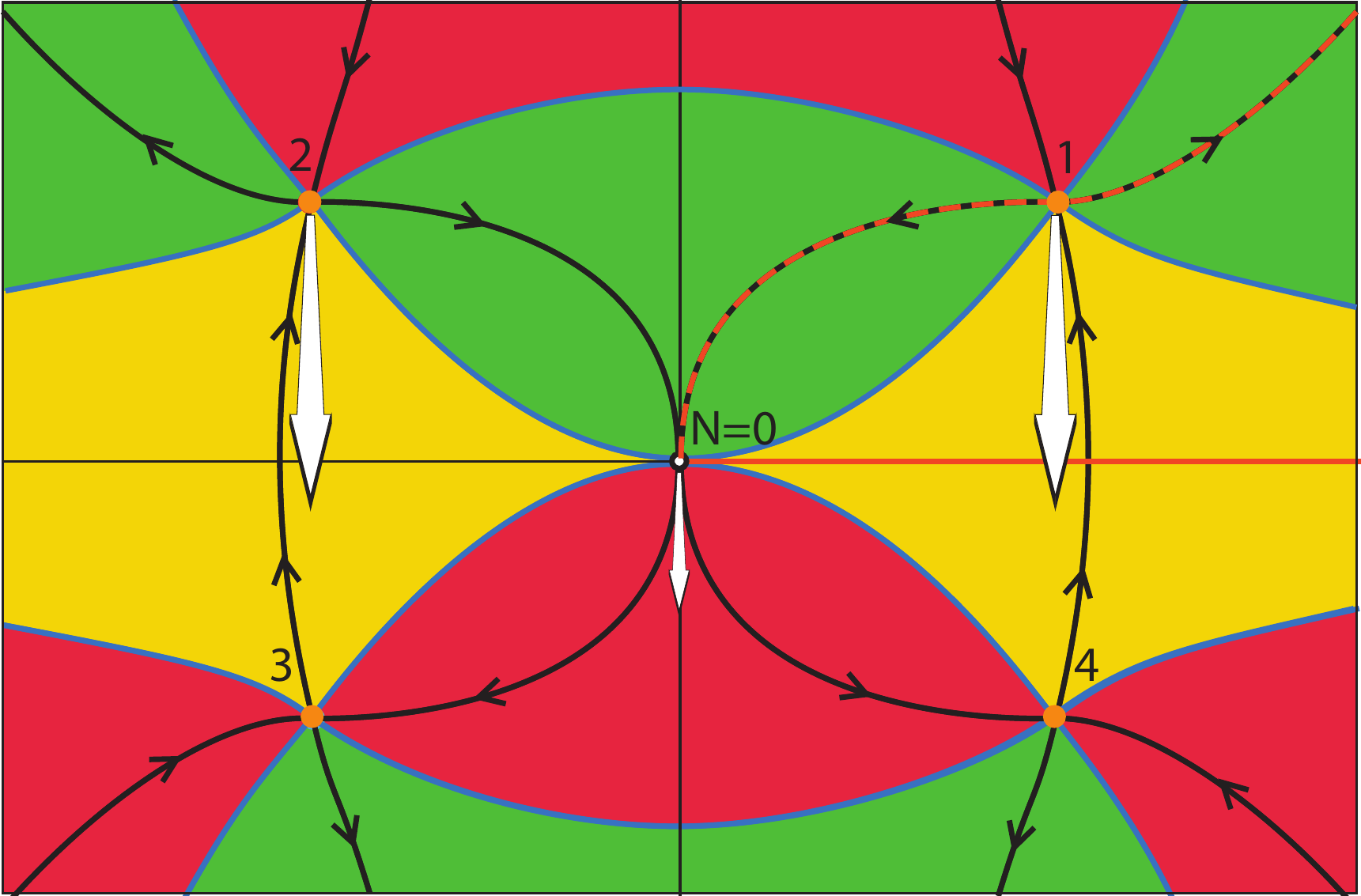}
\caption{The minisuperspace path integral contains $4$ saddle points (orange dots) in the plane of the complex lapse $N$. Steepest descent/ascent lines of the magnitude of the action are drawn as black lines, with the arrows indicating descent. Asymptotic regions of convergence are shown in green, divergent ones are red. For the path integral with Dirichlet boundary conditions, the defining contour of positive real lapse (orange line) can be deformed to the (orange dashed) thimble flowing through saddle point $1$ only. With Robin boundary conditions, saddle points $1,2$ and the singularity of the action at $N=0$ move. Shown here is the direction of motion for negative imaginary $\beta.$} \label{fig:Robin1}
\end{figure}

The difference between the two proposals may be elucidated by considering the saddle points of the full path integral \eqref{prop}. There are four saddle points, with geometries given by $q(t)= H^2 N^2 t (t - 1) + q_1 t,$ with the lapse taking the values $N_{c_1,c_2} = \frac{c_1}{H^2} (\sqrt{H^2 q_1  - 1} + c_2 i)$ for $c_1 , c_2 \in \{ -1 , 1 \}$.  Their action is
\begin{equation}
S(N_{c_1 , c_2}) = c_1 \frac{4\pi^2}{H^2} [ c_2 i - (H^2 q_1 - 1)^{3/2}]\,. \nonumber
\end{equation}
From this expression we see that the propagator \eqref{G} is given by the contribution of the saddle point with $c_1 =1$ and $c_2 = 1,$ in the upper right quadrant of the complex $N$ plane. It is in fact possible to show, applying Picard-Lefschetz theory, that the integral along the positive real $N$ line is equivalent to the integral along the steepest descent path (``thimble'') running through this saddle point alone, see Fig. \ref{fig:Robin1}. This saddle point  is however unstable \cite{FLT2,Halliwell:1989dy}. Thus the result \eqref{G}, though mathematically correct, cannot describe our universe on physical grounds. 

The no-boundary result \eqref{hh} instead would have been obtained by considering the contribution of the two saddle points in the lower half complex $N$ plane, with $c_1 = 1$, $c_2= -1$ and $c_1 = -1$, $c_2= 1$. There is however no convergent contour which can be deformed into a steepest descent path running through solely these two saddle points \cite{FLT3}. In this sense, the wavefunction \eqref{hh} is not the saddle point approximation of the no boundary wavefunction for any Lorentzian path integral with these boundary conditions. With different boundary conditions, the situation may change, as we will now discuss.


Let us consider augmenting the action with a Robin boundary term at the initial surface,
\begin{equation}
S_{tot}= S + \alpha q_0 + \frac{q_0^2}{2 \beta} \,. \nonumber
\end{equation}
The variation of the full action is now
\begin{align}
\delta S_{tot} = & 2\pi^2 \int_0^1 dt \Bigl[ \frac{3 \ddot{q}}{2 N} - 3 N H^2 \Bigr] \, \delta q  - \frac{3 \pi^2}{N} \dot{q}_1 \, \delta q_1 + \mathcal{B} \,\delta q_0\,. \nonumber
\end{align}
Thus we can see that the variational principle is well defined if we impose $ \delta q_1 = 0$ and
\begin{align}
\mathcal{B} \equiv \frac{3 \pi^2}{N} \dot{q}_0 + \alpha + \frac{q_0}{\beta}  = 0 \,. \nonumber
\end{align}
In other words we fix the field value at $t=1$ to be $q(t=1) = q_1$ corresponding to a Dirichlet boundary condition, while at $t= 0$ we impose a condition on the linear combination of $q_0$ and $\dot{q}_0$. With these boundary conditions, the solution of the equation of motion reads
\begin{align}
q(t) = &H^2 N^2 t^2 - \frac{(\alpha \beta + q_1 - H^2 N^2)}{3 \pi^2 \beta - N} N t \nonumber \\ & + \frac{3\pi^2 (q_1 - H^2 N^2) + N \alpha}{3\pi^2 \beta  - N} \beta \label{qsol}
\end{align}
Plugging this solution back into the action we find that the saddle points are
\begin{equation}
N_s = 3 \pi^2 \beta + c_1 \frac{\sqrt{H^2 q_1 -1}}{H^2} + c_1 c_2 \frac{ \sqrt{9 \pi^4 \beta H^4  -   \alpha \beta H^2  -1}}{H^2} \nonumber
\end{equation}
A crucial requirement in order to obtain an implementation of the no-boundary idea is that at the saddle point, the geometry should be of Hawking type, and in particular it should start at zero size. From \eqref{qsol} one can see that the initial size $\bar{q}_0$ vanishes at one (or more) of the saddle points if $\alpha = \pm 6\pi^2 i$ or $\beta = 0 $. 
For $\alpha = + 6\pi^2 i$, this occurs at the ``tunnelling'' saddle points $c_1=c_2=1$ and $c_1=c_2=-1.$ These are however unstable, so that in the following we will consider $\alpha = - 6\pi^2 i$ (we will discuss the meaning of $\beta$ momentarily). This gives $\overline{q}_0=0$ for
\begin{equation}
N_{3,4} = \frac{- i }{H^2} \mp \frac{ \sqrt{H^2 q_1 - 1}}{H^2}\,, \nonumber
\end{equation}
which are precisely the Hartle-Hawking saddle points. The other two saddle points are now located at
\begin{equation}
N_{1,2} =  \frac{i }{H^2} \pm \frac{ \sqrt{H^2 q_1 - 1}}{H^2} + 6\pi^2 \beta \nonumber
\end{equation}
and their initial size is $\overline{q}_0 = 12\pi^2 \beta (i + 3 \pi^2 \beta H^2)$. Therefore requiring that at least one of the saddle point geometries starts out at zero size corresponds to a specific value of $\alpha$ but leaves $\beta$ free.   The value of $\beta$ in fact determines which saddle point(s) are relevant to the path integral. 

For $\beta = 0 $, the relevant saddle point is $N_1$. In this limit the Robin boundary condition reduces to the Dirichlet boundary value $q_0 = 0 $. For non-zero $\beta$ the saddle points in the upper half plane move, and the singularity in the action, which originally resides at $N^\star=0,$ is also shifted to $N^\star= 3 \pi^2 \beta$. Importantly, the lower saddle points (numbers $3$ and $4$ in Fig. \ref{fig:Robin1}), which correspond to the desired Hartle-Hawking geometries, stay put. 

Our strategy will be to define the path integral on a thimble, for the simple reason that it is then manifestly convergent. This has the important consequence that the partial integrations over the scale factor, the lapse and any other fields that might be present, can be performed in any desired order without changing the end result (i.e. Fubini's theorem applies). At vanishing $\beta$ our defining integration contour is simply the Lorentzian one, along real positive values of $N.$ This can then be deformed, using Picard-Lefschetz theory, to the thimble passing through the unstable saddle point $1,$ as shown in Fig. \ref{fig:Robin1}. As $\beta$ is turned on, the singularity of the integral at $N=0$ shifts to $N^\star,$ and thus, in order to maintain an invariant definition, we will consider as our integration contour the thimble(s) emanating from $N^\star.$ 

\begin{figure}
\includegraphics[scale=0.45]{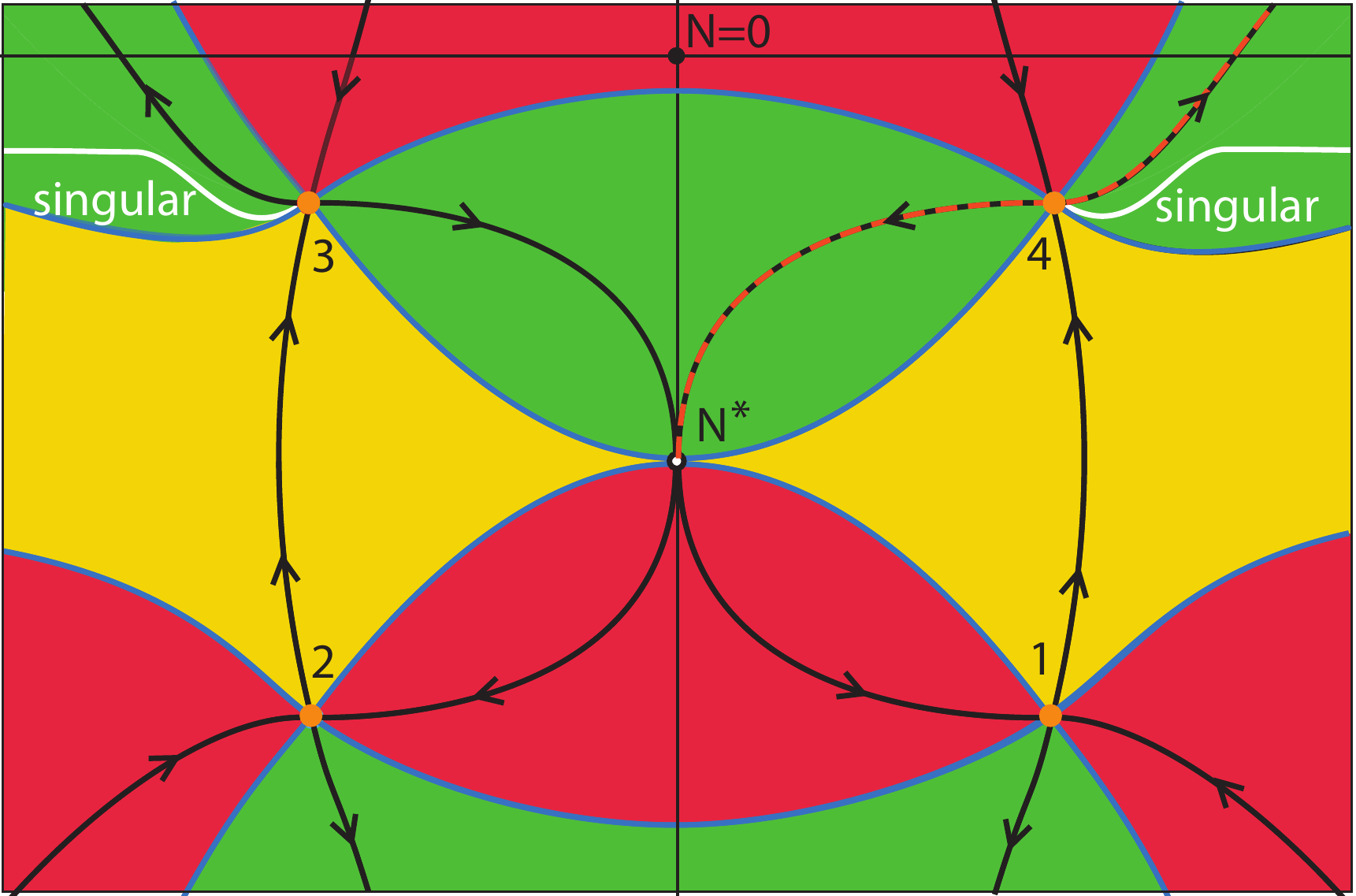}
\caption{For sufficiently large negative imaginary $\beta,$ the unstable saddle point $1$ moves below the Hartle-Hawking saddle point $4$. The thimble emanating from the singularity of the action at $N^\star$ now only passes through this stable saddle point, and the problem with instabilities is avoided. In white we indicate the locus of geometries that contain a singularity, and which we require the thimble to avoid.} \label{fig:Robin2}
\end{figure}

But which value should $\beta$ take? Roughly speaking, the unstable saddle points remain relevant until $\beta$ is large enough in magnitude so that they have moved ``out of the way''. For real negative $\beta$ this approximately means that they have to move further to the left than $N^\star,$ which marks the origin of the integration contour. But since the real part of the saddle points depends on $q_1,$ this is non-sensical from a physical point of view, as the initial conditions would have to keep being readjusted as the universe keeps expanding! On the other hand, in the imaginary $N$ direction, the unstable point (with an original location at $Im(N_1)=+i/H^2$) only has to move beyond the stable one at $Im(N_4)=-i/H^2.$ This condition is independent of the final size of the universe, and thus we will only consider imaginary $\beta$. The minimal magnitude of $\beta$ is determined precisely by the condition that the unstable saddle point move below the stable one,
\begin{align}  
|\beta| > \beta_{min} = \frac{1}{3 \pi^2 H^2}\,. \nonumber
\end{align}
For $\beta$ being negative imaginary and of magnitude larger than this minimal value, the Hartle-Hawking saddle point becomes the only relevant one -- see Fig. \ref{fig:Robin2} for an illustration. Thus we have successfully isolated the Hartle-Hawking saddle point, leading to the propagator
 \begin{equation}
 G[q_1 , 0]_{Robin} = e^{+ \frac{4\pi^2}{H^2 \hbar} - i \frac{4\pi^2 H}{\hbar} (q_1 - \frac{1}{H^2})^{3/2}}\,. \nonumber
\end{equation} 
In fact, we also have the possibility of including both thimbles emanating from $N^\star,$ passing through saddles $3$ and $4$ in Fig. \ref{fig:Robin2}. If we define the integration contour to run over all real values of $N$, from minus infinity to plus infinity, the result will be the real no-boundary wavefunction \eqref{hh} proposed by Hartle and Hawking. Moreover, in the latter case the wave function will satisfy the homogeneous Wheeler-DeWitt equation $\hat{\cal H} \Psi = 0$ (cf. the appendix of Ref. \cite{FLT1}).



\begin{figure}
\includegraphics[scale=0.45]{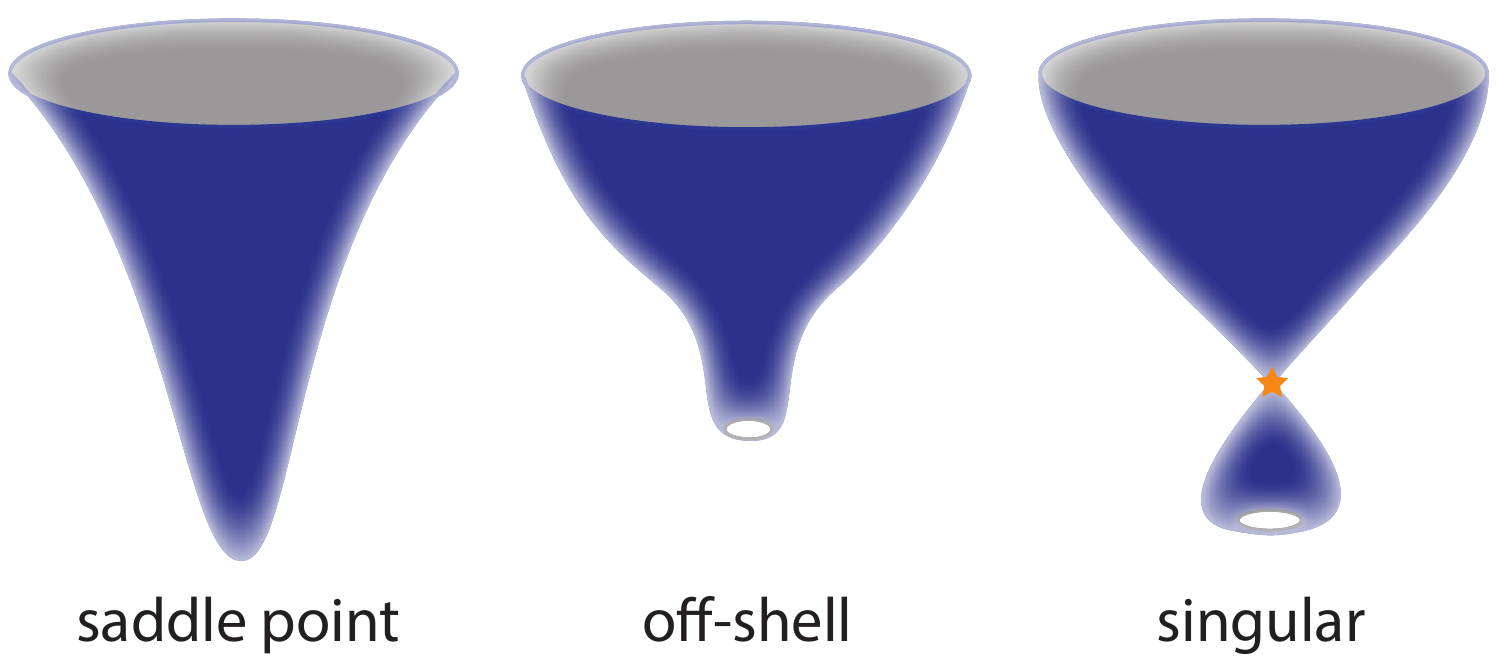}
\caption{On the left we have the smooth saddle point geometry of Hartle-Hawking type. By contrast, with Robin boundary conditions a typical off-shell geometry will not start at zero size (middle). Some off-shell geometries contain a recollapse to zero size, and it is these geometries that we would like to avoid summing over (right).} \label{fig:regsing}
\end{figure}

The Robin boundary condition implies a relationship between the initial size and the initial momentum of the geometries summed over in the path integral. It does however not necessarily avoid the appearance of singularities, in the sense that it may be possible that some of the off-shell geometries contain a region where the size of the universe passes through zero, see Fig. \ref{fig:regsing}. We would like to avoid summing over such geometries, as they are (infinitely) sensitive to the addition of higher curvature terms. This will imply an upper bound on $|\beta|.$

Thus we must analyse the locus of geometries containing a region of zero volume. We start by noting that the imaginary part of the scale factor vanishes at
\begin{equation}
\tau = - \frac{3\pi^2 i \beta}{ H^2 m } \frac{[-q_1 + 6\pi^2 i \beta + H^2(n^2 + m^2 + 6\pi^2 i \beta m )]}{[n^2 + (m + 3 \pi^2 i \beta)^2 ]}\,, \nonumber
\end{equation}
where we have split the lapse into real and imaginary parts as $N = n + i \, m $.  We must then determine if the real part of the scale factor \eqref{qsol} may simultaneously vanish, with $0 \leq \tau \leq 1$.    There is no concise analytic expression for this function. At large magnitudes of the lapse, this singular curve quickly becomes horizontal, see Fig. \ref{fig:Robin2} for a sketch. Therefore the relevant question is whether near the saddle point the thimble, which asymptotically runs off to infinity at an angle of $\pi/6$ \cite{FLT1}, can avoid crossing the singular curve. Thus at $N_4$ we want the angle $\theta$ of the thimble to be larger than the angle $\varphi$ of the singular curve. This latter angle is given by
\begin{equation}
\tan(\varphi) = \frac{(- 2 + 3\pi^2  i \beta H^2) \sqrt{H^2 q_1 - 1}}{(q_1 + 3\pi^2 i \beta )H^2 - 4}\,.\nonumber
\end{equation} 
Meanwhile, the angle $\theta$ is found to be
\begin{equation}
\tan(\theta) = \frac{-1 + 3\pi^2 i \beta H^2}{\sqrt{H^2 q_1  - 1} + H \sqrt{q_1 - 6 \pi^2 i \beta  - 9 \pi^4  H^2 \beta^2}}\,, \nonumber
\end{equation}
where we have assumed that $|\beta|> \beta_{min}$. The condition $\theta > \varphi$ is then satisfied for negative imaginary $\beta$ with
\begin{equation}
|\beta| < \beta_{max} = \frac{{1}}{{3 \pi^2 H^2}}\frac{{ 3 H^2 q_1 -4}}{{ H^2 q_1 -2}} \approx \frac{1}{\pi^2 H^2}\,,\nonumber
\end{equation}
where we assumed $q_1 > \frac{2}{H^2}.$ Thus for negative imaginary $\beta$ with magnitude between $\beta_{min}$ and $\beta_{max}$ the path integral is well defined and contains only the Hartle-Hawking saddle point(s).

Let us now explicitly verify that the perturbations are indeed suppressed. In our model we only have gravitational wave perturbations to deal with. For a single mode with fixed polarisation, the action at quadratic order is given by (see e.g. \cite{FLT2,FLT3})
\begin{equation}
S^{(2)} = \pi^2 \int_0^1 dt \, N \Bigl[q^2 \frac{\dot{\phi}^2}{N^2} - l (l + 2) \phi^2 \Bigr] \nonumber
\end{equation} 
where $\phi$ denotes the magnitude of the perturbation, which has been expanded in spherical harmonics with $l\geq 2$ being the principal quantum number. The extension to a sum over all modes is straightforward. The equation of motion for $\phi$ is thus
\begin{equation}
\ddot{\phi} + 2 \frac{\dot{q}}{q} \dot{\phi} + \frac{N^2 }{q^2} l (l + 2) \phi = 0\,.  \nonumber
\end{equation}
In order to solve it, it is convenient to rewrite the scale factor as $q(t)= H^2 N^2 (t - \gamma)(t - \delta),$ where $\gamma, \delta$ can be read off from \eqref{qsol}. Then we find that two linearly independent solutions for $\phi(t)$ are $f(t)/\sqrt{q}$ and $g(t)/\sqrt{q}$ with
\begin{align}
f(t),g(t) = [\frac{t - \delta}{t - \gamma}]^{\pm \frac{\mu}{2}} [(1 \mp \mu)(\gamma - \delta)+ 2 (t - \gamma)] \nonumber 
\end{align}
and $\mu^2 = 1 - 4l(l+2)/[(\gamma-\delta)^2 N^2H^4]\,.$ At the saddle point the parameters reduce to $\mu (N_4) = (l + 1), \, \gamma (N_4) = 0, \, \delta (N_4) = - 2 i/(\sqrt{H^2 q_1 -1} -i)\,.$ It is straightforward to check that this implies that the mode $f$ blows up at $t=0$ whereas $g/\sqrt{q} |_{t=0} = 0 $. Thus regularity selects the mode $g,$ whose action at the saddle point is given by
\begin{align}
S^{(2)} = i \, \phi_1^2 \frac{l (l + 1) (l +2) }{2 H^2} - \phi_1^2 \frac{l (l + 2) \sqrt{q_1}}{2 H} + \mathcal{O} \Bigl( \frac{1}{\sqrt{q_1}} \Bigr)\,, \nonumber
\end{align}
where $\phi_1$ is the final (real) magnitude of the perturbation. The resulting amplitude $e^{i S^{(2)}/\hbar} $ describes Gaussian distributed perturbations with a scale-invariant spectrum. We conclude therefore that the model is stable against small deviations since large values of $\phi_1$ are suppressed.  

More generally, we can specify Robin boundary conditions for the perturbations.
The full action for the perturbations is then $S_{tot}^{(2)} =S^{(2)} + \alpha_{\phi} \phi_0 + \frac{\phi_0^2}{2 \beta_{\phi}}$ which leads to $\delta \phi_1 = 0$ and
\begin{align}
\delta \phi_0 [- \frac{2\pi^2 q_0^2 \dot{\phi}_0}{N^2} + \alpha_{\phi} +  \frac{\phi_0}{\beta_\phi} ] = 0\,. \nonumber
\end{align}
Any Robin boundary condition with $\alpha_\phi = 0$ and arbitrary $\beta_\phi$ will then retain the stable mode at the saddle point, while also specifying initial conditions for the perturbations off-shell. Since we are only summing over non-singular geometries off-shell, for sufficiently small perturbation amplitudes the approximation of linear perturbation theory around the background will be justified, as the perturbations encounter no divergence anywhere. Thus, the inclusion of perturbations will leave the path integral well-defined, and stable.



As already pointed out in the case of perturbations by Vilenkin and Yamada \cite{VY1,VY2}, the Robin boundary condition can also be implemented as a Gaussian integral 
\begin{align}
\int \, dN \, dq \, e^{iS/\hbar} \int \, dq_0 \, e^{i\alpha q_0/\hbar - \frac{q_0^2}{2|\beta|}} \,,\nonumber
\end{align}
where we must include an integration over the initial scale factor $q_0.$ Owing to the fact that $\beta$ has to be negative imaginary, this may then also be interpreted as an initial coherent state, albeit one with a Euclidean momentum $\alpha$. The presence of a Euclidean momentum not only implements the idea of closing the geometry off in Euclidean time, but it also adds a positive weighting to the associated geometries. In this way we can obtain a final result with an enhanced weighting $e^{+4\pi^2/(\hbar H^2)}$, which in the implementation with Dirichlet boundary conditions was simply impossible \cite{FLT3}. Note that in this context $\sqrt{|\beta|}$ takes on the role of the uncertainty in the initial size $q_0.$ And because we have a coherent state, the uncertainty in the initial momentum is then simply its inverse,
\begin{align}
\Delta q_0 = \sqrt{|\beta|} \sim \frac{1}{H}\,, \quad \Delta p_0 = \frac{\hbar}{\sqrt{|\beta|}} \sim \hbar H\,.\nonumber
\end{align}
Thus we see that in order to have a well defined path integral, the uncertainty must be shared between the initial size and the initial momentum, with the uncertainty being of order the Hubble length.

In concluding, let us note that the present implementation of the no-boundary proposal effectively corresponds to a redefinition. Originally, it was formulated as a sum over compact and regular geometries \cite{HH}. Here the sum is redefined to be over geometries with approximately zero initial size and approximately Euclidean initial momentum. This is just fuzzy enough to allow quantum theory to pick out a regular no-boundary geometry at the saddle point, but the uncertainty is not so large that singular geometries start contributing too.

\noindent {\it Acknowledgments:} We would like to thank Sebastian Bramberger, Job Feldbrugge and Neil Turok for numerous discussions. We gratefully acknowledge the support of the European Research Council in the form of the ERC Consolidator Grant CoG 772295 ``Qosmology''.

\bibliographystyle{utphys}
\bibliography{Robin}

\end{document}